\documentclass[11pt]{article}
\usepackage{amsmath,amsfonts,amssymb}
\newcommand{\sfrac}[2]{{\textstyle{\frac{#1}{#2}}}}

\begin{document}
\newpage
\pagestyle{empty}

\vfill

\rightline{DSF-20/03}
\rightline{LAPTH-987/03}
\rightline{cond-mat/0306636}

\vfill

\begin{center}

{\Large \textbf{\textsf{Sum rules for free energy and frequency 
distribution of \\[3mm] DNA dinucleotides }}}

\vspace{10mm}

{\large L. Frappat$^{ac}$, A. Sciarrino$^{b}$}

\vspace{10mm}

\emph{$^a$ Laboratoire d'Annecy-le-Vieux de Physique Th{\'e}orique LAPTH,}

\emph{CNRS, UMR 5108, and Universit{\'e} de Savoie,}

\emph{BP 110, F-74941 Annecy-le-Vieux Cedex, France}

\vspace{7mm}

\emph{$^b$ Dipartimento di Scienze Fisiche, Universit{\`a} di Napoli
``Federico II''}

\emph{and I.N.F.N., Sezione di Napoli,}

\emph{Complesso Universitario di Monte S. Angelo,}

\emph{Via Cintia, I-80126 Naples, Italy}

\vspace{7mm}
 
\emph{$^c$ Member of Institut Universitaire de France}

\vspace{12mm}

\end{center}

\vspace{7mm}

\begin{abstract}
The large discrepancy between the values of the free energy for DNA 
dinucleotides (or dimers) measured by different teams has raised a yet 
unsettled debate. Here the free energy is fitted by a three parameter 
empiric formula derived in the framework of the crystal basis model of 
genetic code. Sum rules are derived and compared satisfactorily with the 
data. On the basis of theoretical and phenomenological arguments, a 
relation between the correlation functions of dimer distribution and the 
free energy is assumed. From consistency conditions, sum rules are derived. 
A check of these conditions with different samples of experimental data is 
performed, allowing us to argue on the reliability of the different sets of 
experimental data.
\end{abstract}

\vfill
\vfill
\vspace*{3mm}

\hrule

\vspace*{3mm}

\noindent 
Corresponding author: \texttt{frappat@lapp.in2p3.fr}.

\newpage
\pagestyle{plain}
\setcounter{page}{1}
\baselineskip=16pt
\parindent=0pt

\section{Introduction}

The importance of the computation of the free energy $\Delta G^{0}$ and 
enthalpy $ \Delta H^0$ for DNA dinucleotides or dimers was recognized in 
the eighties by many authors and several experimental measures have been 
performed. The experimental values however range in an unacceptable wide 
range. A few years ago SantaLucia \cite{Santa} has performed an accurate 
analysis and comparison of the data from seven laboratories (see Table 
\ref{table:santa} taken from ref. \cite{Santa}, where we have replaced the 
original values of the column Benight \cite{benight2} with the more recent 
ones \cite{benight}), reaching the conclusion that six of the studies were 
actually in agreement and providing explanations for the discrepancies. In 
an attempt to settle by thermodynamics arguments the controversy, 
Miramontes and Cocho \cite{MC} have analysed quite recently the same set of 
data by assuming a relation between the correlation function of the dimers 
and their free energy, reaching the conclusion that the most reliable set 
of values is just the one which was excluded by SantaLucia. Indeed in ref. 
\cite{MC} a linear relation between the correlation function for the dimer 
and the corresponding free energy was postulated, which allowed these 
authors to determine which set of experimental data was in better agreement 
with the postulated relation. A shortcoming of this analysis is that the 
sum of the free energies for strong dimers does not satisfy an identity 
derived from the postulated equation. The purpose of this work is to come 
back to this controversial question. First, we propose a theoretical 
formula to compute the free energy, from which sum rules are derived and 
compared with the values of experimental data. Second, we motivate the 
assumption of a relation between the correlation function and the free 
energy, different from the one assumed in \cite{MC}, which satisfies 
trivial identities required by the definition of the correlation functions. 
We make several consistency checks and we try to determine the reliability 
of the experimental values, comparing with the calculated values of the 
correlation matrix in \cite{MC}.

\section{Fit for the free energy}

Let us recall that a mathematical framework was proposed \cite{FSS1}, in 
which the codons appear as composite states of nucleotides. The four 
nucleotides being assigned to the fundamental irreducible representation of 
the quantum group ${\cal U}_{q}(sl_{H}(2) \oplus sl_{V}(2))$ in the limit 
$q \to 0$ (the indices $H$ and $V$ distinguish the two $sl(2)$), a sequence 
of $N$ nucleotides is described by a pure state in the $N$-fold tensor 
product of the fundamental representation. In particular, dimers or 
dinucleotides are obtained as the two-fold tensor product, the labels 
specifying the irreducible representation to which they belong being given 
in Table \ref{table:dimer}. In ref. \cite{FSS1} we have fitted old 
experimental data of the free energy $\Delta G^0_{37}$ (for simplicity we 
will omit the temperature label in the following) for RNA dinucleotides 
with a 4 parameter formula built up with the generators of ${\cal U}_{q \to 
0}(sl_{H}(2) \oplus sl_{V}(2))$ and in \cite{FSS2} the more recent data of 
\cite{ST} have been fitted with the following 2 parameter formula
\begin{equation}
\Delta G^{0}  = \alpha  + \beta \, (C_{H} + C_{V}) J_{3H} 
\label{eq:dg}
\end{equation}
where $J_{3X}$ ($X = H$ or $V$) stands for the diagonalized $sl(2)_{X}$ 
generator and $C_{X}$ is the Casimir operator of ${\cal U}_{q \to 
0}(sl(2)_{X})$ for the considered dimer $ij$. Let us recall that the 
Casimir operator eigenvalue in the $J$-representation is $J(J+1)$. In order 
not to overload the notation, here and in the following, we will not 
explicitly write the labels of the dimer, if not necessary to identify a 
specific dimer.

\medskip

Here we propose for the DNA dinucleotides a 3 parameter formula, which is a 
generalisation of eq. (\ref{eq:dg}):
\begin{equation}
\Delta G^0 = \alpha_{0} + \alpha_{1} \, J_{3H} + \alpha_{2} \, 
(J_{3V})^{2} \, (2J_{3H} + 1)
\label{eq:fe}
\end{equation}
This equation leads to the following theoretical values of the dimer free 
energies $\Delta G^{0}$ in terms of the parameters $\alpha_{0}$, 
$\alpha_{1}$, $\alpha_{2}$:
\begin{center}
\begin{tabular}{|lc|lc|}
\hline
AA/TT & $\alpha_{0} - \alpha_{1} - \alpha_{2}$ & CT/GA & $\alpha_{0} + 
\alpha_{2}$ \\
AT/TA & $\alpha_{0} - \alpha_{1}$ & GA/CT & $\alpha_{0} + \alpha_{2}$ \\
TA/AT & $\alpha_{0} - \alpha_{1}$ & CG/GC & $\alpha_{0} + \alpha_{1}$ \\
CA/GT & $\alpha_{0}$ & GC/CG & $\alpha_{0} + \alpha_{1}$ \\
GT/CA & $\alpha_{0}$ & GG/CC & $\alpha_{0} + \alpha_{1} + 3\alpha_{2}$ \\
\hline
\end{tabular}
\end{center}
A best-fit procedure allows one to evaluate these parameters as follows: 
\begin{equation}
\label{eq:param}
\alpha_{0} = \sfrac{1}{116} (14N_{1} + 4N_{2} - 6N_{3}) \;, \quad
\alpha_{1} = \sfrac{1}{116} (4N_{1} + 26N_{2} - 10N_{3}) \;, \quad
\alpha_{2} = \sfrac{1}{116} (-6N_{1} - 10N_{2} + 15N_{3})
\end{equation}
where (we specify by a couple of indices the free energy of a dinucleotide)
\begin{eqnarray}
N_{1} &=& \Delta G^{0}_{GG} + \Delta G^{0}_{CG} + \Delta G^{0}_{GC} + 
\Delta G^{0}_{CT} + \Delta G^{0}_{GA} + \Delta G^{0}_{GT} + \Delta 
G^{0}_{CA} + \Delta G^{0}_{TA} + \Delta G^{0}_{AT} + \Delta G^{0}_{AA} 
\nonumber \\
N_{2} &=& \Delta G^{0}_{GG} + \Delta G^{0}_{GC} + \Delta G^{0}_{CG} - 
\Delta G^{0}_{AA} - \Delta G^{0}_{AT} - \Delta G^{0}_{TA} \nonumber \\
N_{3} &=& 3\Delta G^{0}_{GG} + \Delta G^{0}_{CT} + \Delta G^{0}_{GA} - 
\Delta G^{0}_{AA}
\end{eqnarray}
Hence we get for the different studies, see Table \ref{table:santa}, the 
best-fit values of the parameters $\alpha_{0}$, $\alpha_{1}$, $\alpha_{2}$:
\begin{center}
\begin{tabular}{|l|cccccccc|}
\hline
& Gotoh & Vologodskii & Breslauer & Delcourt & SantaLucia & Sugimoto & 
Unified & Benight \\
& \cite{gotoh} & \cite{volog} & \cite{breslauer} & \cite{delcourt} & 
\cite{santal} & \cite{sugim} & \cite{unified} & \cite{benight} \\
\hline
$\phantom{-}\alpha_0$ & 0.98 & 1.37 & 1.89 & 1.24 & 1.53 & 1.71 & 1.47 & 
1.35 \\
$\phantom{-}\alpha_1$ & 0.70 & 0.60 & 0.99 & 0.61 & 0.66 & 0.81 & 0.73 & 
0.54 \\
$-\alpha_2$ & 0.14 & 0.12 & 0.18 & 0.09 & 0.15 & 0.16 & 0.14 & 0.03 \\
$\phantom{-}s^{2}$ & 0.0015 & 0.0011 & 0.1577 & 0.0014 & 0.0114 & 0.0199 & 
0.0070 & 0.0069 \\
$\phantom{-}\chi^{2}$ & 0.0243 & 0.0099 & 1.0001 & 0.0167 & 0.0753 & 0.0992 
& 0.0821 & 0.0590 \\
\hline
\end{tabular}
\end{center}
The last two rows correspond to the square mean deviation $s^{2} = 
\frac{1}{N} \sum (y_{exp}-y_{th})^{2}$ ($N$ is the number of points, here 
10) and to $\chi^{2} = \sum (y_{exp}-y_{th})^{2}/y_{th}$. Evaluation of the 
incomplete Gamma function, which is an estimate of the goodness-of-fit, 
shows that the fit is good with a confidence level greater than 95\%. Table 
\ref{table:fitdeltag} gives the fitted absolute values for dimer free 
energy parameters $\Delta G^0$ corresponding to the different samples. From 
an inspection of the values of $s^{2}$ and $\chi^{2}$, one sees that eq. 
(\ref{eq:fe}) is well fitted by the different sets of experimental data, 
except by the ones from Breslauer.

\section{Sum rules}

We derive from eq. (\ref{eq:fe}) a set of identities and sum rules. First, 
it is clear that
\begin{equation}
\Delta G^0_{ij} = \Delta G^0_{ji} \qquad \mbox{and} \qquad \sum_{j=A,C,G,T} 
\, \Delta G^0_{ij} \; = \sum_{j=A,C,G,T} \, \Delta G^0_{ji}
\label{eq:simm}
\end{equation}
In particular we get
\begin{eqnarray}
\sum_{j=A,C,G,T} \, \Delta G^0_{Cj} &=& \sum_{j=A,C,G,T} \, \Delta 
G^0_{Gj} \;\; = \;\; 4 \alpha_{0} + 2 \alpha_{1} + 4 \alpha_{2} 
\label{eq:sommes} \\
\sum_{j=A,C,G,T} \, \Delta G^0_{Aj} &=& \sum_{j=A,C,G,T} \, \Delta 
G^0_{Tj} \;\; = \;\; 4 \alpha_{0} - 2 \alpha_{1} 
\label{eq:sommew} \\
\sum_{i,j=A,C,G,T} \, \Delta G^0_{ij} &=& 16 \, \alpha_{0} + 8 \, 
\alpha_{2}
\label{eq:sommet}
\end{eqnarray}
In Table \ref{table:sommeexp} we report the experimental values computed 
using the values of Table \ref{table:santa}. Note that in ref. \cite{MC} 
the existence of the sum rules eqs. (\ref{eq:sommes}) and (\ref{eq:sommew}) 
was already remarked, but the two equations should have the same values, 
which is actually not the case.
\begin{table}[htbp]
\centering
\caption{$\Big.$ Experimental values of the sums of free energies [see eq. 
(\ref{eq:simm})].
\label{table:sommeexp}}
\begin{tabular}{|l|cccccccc|}
\hline
& Gotoh & Vologodskii & Breslauer & Delcourt & SantaLucia & Sugimoto & 
Unified & Benight \\
\hline
$\Big. \sum_{i} \Delta G^{0}_{Ci}$ & 4.72 & 6.16 & 9.18 & 5.78 & 6.72 & 
8.10 & 6.74 & 6.54 \\
$\Big. \sum_{i} \Delta G^{0}_{Gi}$ & 4.77 & 6.20 & 8.11 & 5.80 & 6.94 & 
7.40 & 6.82 & 6.26 \\
$\Big. \sum_{i} \Delta G^{0}_{Ti}$ & 2.55 & 4.27 & 5.63 & 3.68 & 5.08 & 
5.30 & 4.33 & 4.45 \\
$\Big. \sum_{i} \Delta G^{0}_{Ai}$ & 2.51 & 4.21 & 5.33 & 3.74 & 4.51 & 
5.10 & 4.60 & 4.27 \\
\hline
\end{tabular}
\end{table}

Due to the complementarity rule, one has
\begin{eqnarray}
\sum_{i=A,C,G,T} \, \Delta G^0_{Ci} = \sum_{i=A,C,G,T} \, \Delta G^0_{iG} 
\qquad \mbox{and} \qquad \sum_{i=A,C,G,T} \, \Delta G^0_{Gi} = 
\sum_{i=A,C,G,T} \, \Delta G^0_{iC} \label{eq:sum1} \\
\sum_{i=A,C,G,T} \, \Delta G^0_{Ai} = \sum_{i=A,C,G,T} \, \Delta G^0_{iU} 
\qquad \mbox{and} \qquad \sum_{i=A,C,G,T} \, \Delta G^0_{Ui} = 
\sum_{i=A,C,G,T} \, \Delta G^0_{iA} \label{eq:sum2}
\end{eqnarray}

Now we derive also news sum rules
\begin{eqnarray}
& \Delta G^0_{CG} + \Delta G^0_{TA} \;=\; 2 \, \Delta G^0_{TG} \;=\; 2 \, 
\Delta G^0_{AC} \label{eq:10} \\
& \Delta G^0_{CC} + \Delta G^0_{TT} \;=\; 2 \,\Delta G^0_{TC} \;=\; 2 \, 
\Delta G^0_{GA} \label{eq:11} \\
& \Delta G^0_{CC} + \Delta G^0_{AA} \;=\; 2 \, \Delta G^0_{TC} \;=\; 2 \, 
\Delta G^0_{AG}
\label{eq:11d}
\end{eqnarray}
We report in Table \ref{table:sommeexp2} a comparison with the experimental 
data, making an average of the different experimental values, theoretically 
equal due to eq. (\ref{eq:fe}), i.e.
\begin{eqnarray}
&& S_{1} \;=\; \Delta G^0_{CG} + \Delta G^0_{TA} + \Delta G^0_{GC} + \Delta 
G^0_{AT} - \Delta G^0_{TG} - \Delta G^0_{GT} - \Delta G^0_{AC} - \Delta 
G^0_{CA} \;=\; 0 \label{eq:aver} \\
&& S_{2} \;=\; \Delta G^0_{CC} + \Delta G^0_{TT} + \Delta G^0_{GG} + \Delta 
G^0_{AA} - \Delta G^0_{CT} - \Delta G^0_{TC} - \Delta G^0_{AG} - \Delta 
G^0_{GA} \;=\; 0 \label{eq:aver2}
\end{eqnarray}
\begin{table}[htbp]
\centering
\caption{$\Big.$ Sum rules for free energies [see eqs. 
(\ref{eq:aver})--(\ref{eq:aver2})].
\label{table:sommeexp2}}
\begin{tabular}{|l|cccccccc|}
\hline
& Gotoh & Vologodskii & Breslauer & Delcourt & SantaLucia & Sugimoto & 
Unified & Benight \\
\hline
$S_{1}$ & $-0.07$ & $0.08$ & $2.19$ & $0.10$ & $-0.09$ & $0.50$ & $0.09$ & 
$-0.32$ \\
$S_{2}$ & $-0.22$ & $0.24$ & $3.30$ & $-0.14$ & $0.34$ & $0.60$ & $0.52$ & 
$0.36$ \\
\hline
\end{tabular}
\end{table}

As it can be seen the sum rules are reasonably well satisfied, except for 
the data of Breslauer. However we cannot make any statement on the 
reliability of the different experimental data on the basis of the accuracy 
by which they fit our empirical formula eq. (\ref{eq:fe}).

\section{Dinucleotide distribution}

In order to settle on more theoretical ground our analysis, we consider the 
dimer correlation function. In \cite{MC} the dimer distribution was 
characterized by the correlation function
\begin{equation}
\Gamma_{ij} = f_{ij} \, - \, f_{i}f_{j}
\label{eq:corr}
\end{equation}
where the labels $i,j$ denote the nucleotides, $i,j \in \{A,C,G,T\}$, and 
$f_{i}$ ($f_{ij}$) denote the frequency of the $i$ nucleotide ($ij$ 
dinucleotide). From eq. (\ref{eq:corr}), it follows
\begin{equation}
\sum_{i=A,C,G,T} \, \Gamma_{ij} \;\;=\;\; \sum_{j=A,C,G,T} \, \Gamma_{ij} 
\;\;=\;\; 0
\label{eq:id}
\end{equation}
In \cite{MC} the following relation between $\Gamma_{ij}$ and the free 
energy $\Delta G^0$ was assumed:
\begin{equation}
\Gamma_{ij} = a \, + \, b \, \Delta G^0_{ij}
\label{eq:cf}
\end{equation}
where $a$ and $b$ are biological species dependent parameters. Inserting 
eq. (\ref{eq:fe}) into eq. (\ref{eq:id}) one gets the identity
\begin{equation}
4a \, + \, b \sum_{j=A,C,G,T} \, \Delta G^0_{ij} \;=\; 0 \;\;\; \Rightarrow 
\;\;\; \sum_{j=A,C,G,T} \, \Delta G^0_{ij} \;=\; const. \;\; \mbox{for all 
$i$}
\label{eq:somme}   
\end{equation}
In ref. \cite{MC}, from the data reported in Table \ref{table:santa}, 
except the last column which was not considered, the authors show that eq. 
(\ref{eq:somme}) was satisfied by the weak dimers only, i.e. with label $i 
\in \{A, T \}$. Let us remark: i) that the statistical mechanics motivation 
which led the authors to postulate eq. (\ref{eq:fe}) holds for an isolated 
system, which is not the case for a dimer inserted in a DNA strand; ii) the 
computed values of the correlation matrix, see Table 3 of \cite{MC}, for 
the same biological species, show, in many cases, a much larger variation 
than the corresponding variation of the free energy, changing the $ij$ 
dimer; iii) our empirical formula eq. (\ref{eq:fe}) predicts the dimers 
$ij$ and $ji$ to have the same free energy, which is approximately true 
(see Table \ref{table:santa}), while on the contrary the correlation 
function $\Gamma_{ij}$ is generally non symmetric. From the above remarks 
we assume the following relation between $\Gamma_{ij}$ and $ \Delta 
G^0_{ij}$:
\begin{equation}
\Gamma_{ij} = a + b \, \bigg( \Delta G_{ij}^{0} - \sfrac{1}{4} 
\sum_{k=A,C,G,T} \Big( \Delta G_{ki}^{0} + \Delta G_{jk}^{0} \Big) \bigg) + 
(1 - \delta_{ij}) \; h_{ij}
\label{eq:free}
\end{equation}
where $h_{ij}$ are biological species dependent real coefficients. The 
complementarity implies that the coefficients $h_{ij}$ and 
$h_{\bar\jmath\bar\imath}$ are equal for two complementary dimers $ij$ 
(from $5'$ to $3'$) and $\bar\jmath\bar\imath$ (from $3'$ to $5'$), so 
there is only 8 coefficients $h_{ij}$. \\
The corrective term in the free energy can be considered as a "penalty" due 
to the interaction of the nucleotides of the dimer with the two nearest 
neighbour nucleotides in the strand, assumed uniformly distributed.

\medskip

Since the correlation coefficient $\Gamma_{ij}$ has to satisfy the sum rule 
(\ref{eq:id}) by definition, one is led to the constraints ($\forall j$)
\begin{eqnarray}
0 & = & 4a \;+\; b \, \sum_{i=A,C,G,T} \Big( \Delta G_{ij}^{0} - \Delta 
G_{ji}^{0} \Big) \;-\; \frac{b}{4} \, \sum_{k,i=A,C,G,T} \Delta G_{ki}^{0} 
\;+\; \sum_{i=A,C,G,T} (1 - \delta_{ij}) \; h_{ij} 
\nonumber \\
& = & 4a \;+\; b \, \sum_{i=A,C,G,T} \Big( \Delta G_{ji}^{0} - \Delta 
G_{ij}^{0} \Big) \;-\; \frac{b}{4} \, \sum_{k,i=A,C,G,T} \Delta G_{ik}^{0} 
\;+\; \sum_{i=A,C,G,T} (1 - \delta_{ij}) \; h_{ij}
\label{eq:constr1}
\end{eqnarray}
Eqs. (\ref{eq:simm})--(\ref{eq:sommet}) imply for any pair $(i,j)$ of 
nucleotides
\begin{equation}
2b \; (2\alpha_{0} + \alpha_{2}) \; - \; 4a \;\;=\;\; \sum_{k=A,C,G,T} (1 - 
\delta_{ik}) \; h_{ik} = \sum_{k=A,C,G,T} (1 - \delta_{kj}) \; h_{kj}
\label{eq:constr2}
\end{equation}
As eq. (\ref{eq:constr2}) gives 4 independent relations, we are left with 4 
parameters $ h_{ij}$. We remark that in eq. (\ref{eq:free}) only the 
following combinations of $a$, $b$ and $\alpha_{i}$ parameters appear in 
the free energy term:
\begin{equation}
x = a \, - \, b \, \alpha_{0} \qquad \mbox{and} \qquad y = b \, \alpha_{2}
\label{eq:xy}
\end{equation}
We then deduce from the 4 constraints (\ref{eq:constr2}) the following 
relations among the coefficients $h_{ij}$ (we choose $h_{CA}$, $h_{CT}$, 
$h_{CG}$, $h_{AC}$, $h_{TC}$, $h_{GC}$, $h_{AT}$, $h_{AT}$)
\begin{eqnarray}
&& h_{CG} \, + \, h_{GC} \, - \, h_{AT} \, - \, h_{TA} = 0 \label{eq:c1} \\
&& h_{TC} \, - \, h_{CT} \, + \, h_{GC} \, - \, h_{AT} = 0 \label{eq:c2} \\
&& h_{CA} \, - \, h_{AC} \, + \, h_{CT} \, - \, h_{TC} \, + \, h_{CG} \, - 
\, h_{GC} = 0 \label{eq:c3} 
\end{eqnarray}
Using eq. (\ref{eq:free}) we can replace the following equations by sum 
rules for the corresponding correlation coefficients:
\begin{eqnarray}
&& \Gamma_{CG} \, + \, \Gamma_{GC} \, - \, \Gamma_{AT} \, - \, \Gamma_{TA} 
= -4 y = 2 \, (\Gamma_{AA} \, - \, \Gamma_{CC}) 
\label{eq:corr1} \\
&& \Gamma_{CT} \, - \, \Gamma_{TC} \, + \, \Gamma_{CG} \, - \, \Gamma_{TA} 
= -2 y = \Gamma_{AA} \, - \, \Gamma_{CC} 
\label{eq:corr2} \\
&& \Gamma_{CA} \, - \, \Gamma_{AC} \, + \, \Gamma_{CT} \, - \, \Gamma_{TC} 
\, + \, \Gamma_{CG} \, - \, \Gamma_{GC} = 0 
\label{eq:corr3} 
\end{eqnarray}
The above equations are well satisfied (within $< 5 \%$) by the 
experimental data, see Table 3 of \cite{MC}, therefore we conclude that our 
parametrization (\ref{eq:free}) for the correlation function is 
satisfactory and we can carry on our analysis. \\
Consider the following differences of the correlation coefficients: 
$\Gamma_{CT}-\Gamma_{TC}$, $\Gamma_{TT}-\Gamma_{CC}$ and 
$\Gamma_{AT}-\Gamma_{GC}$. Inserting the theoretical expression 
(\ref{eq:free}) of $\Gamma_{ij}$, one gets for each of the three 
differences:
\begin{eqnarray}
&& \Gamma_{CT}-\Gamma_{TC} = Z_{CT-TC} \; b \; + \; h_{CT} - h_{TC} \\
&& \Gamma_{TT}-\Gamma_{CC} = Z_{TT-CC} \; b \; + \; h_{TT} - h_{CC} \\
&& \Gamma_{AT}-\Gamma_{GC} = Z_{AT-GC} \; b \; + \; h_{AT} - h_{GC}
\label{eq:coefp}
\end{eqnarray}
where the coefficients $Z$ are functions of the free energies $\Delta 
G^{0}$. Summing up the three above equations, one gets that the l.h.s. is 
vanishing, due to eq. (\ref{eq:id}) and the equality of the correlation 
coefficients for complementary dimers, which implies, using eq. 
(\ref{eq:c2}), that the coefficients $Z$ are related:
\begin{equation}
Z_{CT-TC} \, + \, Z_{TT-CC} \, + \, Z_{AT-GC} = 0
\label{eq:relp}
\end{equation}
Let us emphasize that this relation is biological species independent, by 
virtue of eq. (\ref{eq:c2}) valid for each biological species, and by the 
complementarity rule for $\Gamma_{ij}$. \\
Note also that relation (\ref{eq:relp}) is automatically satisfied when 
plugging the theoretical expressions of the free energies of the dimers 
(i.e. in terms of the parameters $\alpha_{0}$, $\alpha_{1}$ and 
$\alpha_{2}$). \\
Analogously using eq. (\ref{eq:c3}) and the complementarity rule we get
\begin{equation}
Z_{CA-GT} + Z_{CT-GA} + Z_{CG-GC} = 0
\label{eq:relp1}
\end{equation}
Note that eq. (\ref{eq:corr1}) is satisfied identically from the 
parametrization  (\ref{eq:free}) and the constraint (\ref{eq:c1}). \\
We report in Table \ref{table:coefp} and Table \ref{table:coefp1} the 
values of the coefficients $Z$ and their sum, calculated with the 
experimental free energies given by the different authors (see table 
\ref{table:santa}). As it can be seen most of the values of the sums are 
quite close to zero, except for Breslauer, SantaLucia and Sugimoto.
\begin{table}[htbp]
\centering
\caption{$\Big.$ Values of the coefficients $Z$ of eq. (\ref{eq:relp}).
\label{table:coefp}}
\begin{tabular}{|l|cccccccc|}
\hline
& Gotoh & Vologodskii & Breslauer & Delcourt & SantaLucia & Sugimoto & 
Unified & Benight \\
\hline
$Z_{CT-TC}$ & $-0.123$ & $-0.115$ & $\phantom{-}0.133$ & $\phantom{-}0.060$ 
& $-0.498$ & $\phantom{-}0.125$ & $\phantom{-}0.027$ & $\phantom{-}0.005$ 
\\
$Z_{TT-CC}$ & $\phantom{-}0.318$ & $\phantom{-}0.220$ & $\phantom{-}0.492$ 
& $\phantom{-}0.160$ & $\phantom{-}0.268$ & $\phantom{-}0.375$ & 
$\phantom{-}0.318$ & $\phantom{-}0.080$ \\
$Z_{AT-GC}$ & $-0.285$ & $-0.205$ & $\phantom{-}0.145$ & $-0.180$ & 
$-0.560$ & $0$ & $-0.155$ & $\phantom{-}0.015$ \\
sum & $-0.090$ & $-0.100$ & $\phantom{-}0.770$ & $\phantom{-}0.040$ & 
$-0.790$ & $\phantom{-}0.500$ & $\phantom{-}0.190$ & $\phantom{-}0.100$ \\
\hline
\end{tabular}
\end{table}

\begin{table}[htbp]
\centering
\caption{$\Big.$ Values of the coefficients $Z$ of eq. (\ref{eq:relp1})
\label{table:coefp1}}
\begin{tabular}{|l|cccccccc|}
\hline
& Gotoh & Vologodskii & Breslauer & Delcourt & SantaLucia & Sugimoto & 
Unified & Benight \\
\hline
$Z_{CA-AC}$ & $-0.013$ & $\phantom{-}0.025$ & $\phantom{-}1.013$ & $-0.110$ 
& $\phantom{-}0.358$ & $\phantom{-}0.425$ & $-0.077$ & $\phantom{-}0.405$ 
\\
$Z_{CT-TC}$ & $-0.123$ & $-0.115$ & $\phantom{-}0.133$ & $\phantom{-}0.060$ 
& $-0.498$ & $\phantom{-}0.125$ & $\phantom{-}0.027$ & $\phantom{-}0.005$ 
\\
$Z_{CG-GC}$ & $\phantom{-}0.035$ & $\phantom{-}0.010$ & $\phantom{-}0.995$ 
& $\phantom{-}0.010$ & $-0.300$ & $\phantom{-}0.850$ & $-0.110$ & 
$\phantom{-}0.150$ \\
sum & $-0.100$ & $-0.080$ & $\phantom{-}2.140$ & $-0.040$ & $-0.440$ & 
$\phantom{-}1.400$ & $-0.160$ & $\phantom{-}0.560$ \\
\hline
\end{tabular}
\end{table}

\section{Conclusions}

We have proposed a 3 parameter formula to fit the free energy for the DNA 
dinucleotides and derived a set of sum rules. We have compared the 
theoretical values with the experimental data of seven authors as well as 
their averaged value. The results of the fits reported in Tables 
\ref{table:fitdeltag} and \ref{table:sommeexp2}, show in the average a 
satisfactory agreement, except for Breslauer. On the basis of the above 
comparison, we cannot make any statement on the reliability of the 
different experimental data. In order to support our analysis by general 
theoretical arguments, we postulate a relation between the free energy and 
the dimer correlation function eq.(\ref{eq:free}), which has theoretical 
motivation from statistical mechanics as well as experimental motivation 
from the analysis of the computed correlation function. Our postulated 
equation satisfies the identity that the sum of correlation functions has 
to satisfy by definition. From consistency equation, we derive a set of sum 
rules for the correlation functions which are well satisfied by the 
computed values for several biological species. This analysis supports the 
validity of our relation eq. (\ref{eq:free}), which allows us to perform 
biological independent consistency checks, which is remarkably verified by 
our theoretical formula. We have checked which set of experimental data 
satisfy the consistency relations. The result is that the data of 
\cite{breslauer}, \cite{santal} and \cite{sugim} are not consistent. 
Therefore we disagree with the conclusions of \cite{MC}. The results of our 
analysis are more close to the ones of \cite{Santa}.

\clearpage

\begin{table}[p]
\centering
\caption{$\Big.$ Dimer representation content.
\label{table:dimer}}
\begin{tabular}{|ccccc||ccccc|}
\hline
dimer & $J_{H}$ & $J_{V}$ & $J_{3H}$ & $J_{3V}$ & dimer & $J_{H}$ & $J_{V}$ 
& $J_{3H}$ & $J_{3V}$ \\
\hline
CC & 1 & 1 & 1 & 1 & GC & 1 & 1 & 1 & 0 \\
CT & 0 & 1 & 0 & 1 & GT & 0 & 1 & 0 & 0 \\
CG & 1 & 0 & 1 & 0 & GG & 1 & 1 & 1 & $-1$ \\
CA & 0 & 0 & 0 & 0 & GA & 0 & 1 & 0 & $-1$ \\
TC & 1 & 1 & 0 & 1 & AC & 1 & 1 & 0 & 0 \\
TT & 1 & 1 & $-1$ & 1 & AT & 1 & 1 & $-1$ & 0 \\
TG & 1 & 0 & 0 & 0 & AG & 1 & 1 & 0 & $-1$ \\
TA & 1 & 0 & $-1$ & 0 & AA & 1 & 1 & $-1$ & $-1$ \\
\hline
\end{tabular}
\end{table}

\begin{table}[p]
\centering
\caption{$\Big.$ Experimental absolute values for dimer free energy 
parameters $\Delta G^0$ (in kcal/mol).  
\label{table:santa}}
\begin{tabular}{|l|cccccccc|}
\hline
& Gotoh & Vologodskii & Breslauer & Delcourt & 
SantaLucia & Sugimoto & Unified & Benight \\
& \cite{gotoh} & \cite{volog} & \cite{breslauer} & \cite{delcourt} & 
\cite{santal} & \cite{sugim} & \cite{unified} & \cite{benight} \\
\hline
AA/TT & 0.43 & 0.89 & 1.66 & 0.67 & 1.02 & 1.20 & 1.00 & 0.91 \\
AT/TA & 0.27 & 0.81 & 1.19 & 0.62 & 0.90 & 0.90 & 0.88 & 0.83 \\
TA/AT & 0.22 & 0.76 & 0.76 & 0.70 & 0.90 & 0.90 & 0.58 & 0.68 \\
CA/GT & 0.97 & 1.37 & 1.80 & 1.19 & 1.70 & 1.70 & 1.45 & 1.54 \\
GT/CA & 0.98 & 1.35 & 1.13 & 1.28 & 1.43 & 1.50 & 1.44 & 1.25 \\
CT/GA & 0.83 & 1.16 & 1.35 & 1.17 & 1.16 & 1.50 & 1.28 & 1.28 \\
GA/CT & 0.93 & 1.25 & 1.41 & 1.12 & 1.46 & 1.50 & 1.30 & 1.30 \\
CG/GC & 1.70 & 1.99 & 3.28 & 1.87 & 2.09 & 2.80 & 2.17 & 1.87 \\
GC/CG & 1.64 & 1.96 & 2.82 & 1.85 & 2.28 & 2.30 & 2.24 & 1.86 \\
GG/CC & 1.22 & 1.64 & 2.75 & 1.55 & 1.77 & 2.10 & 1.84 & 1.85 \\
\hline
\end{tabular}
\end{table}

\begin{table}[p]
\centering
\caption{$\Big.$ Fitted absolute values for dimer free energy parameters 
$\Delta G^0$ (in kcal/mol).
\label{table:fitdeltag}}
\begin{tabular}{|l|cccccccc|}
\hline
& Gotoh & Vologodskii & Breslauer & Delcourt & SantaLucia & Sugimoto & 
Unified & Benight \\
\hline
AA/TT & 0.46 & 0.92 & 1.13 & 0.75 & 1.08 & 1.11 & 0.93 & 0.85 \\
AT/TA & 0.30 & 0.79 & 0.93 & 0.65 & 0.91 & 0.93 & 0.78 & 0.81 \\
TA/AT & 0.30 & 0.79 & 0.93 & 0.65 & 0.91 & 0.93 & 0.78 & 0.81 \\
CA/GT & 1.02 & 1.40 & 1.94 & 1.26 & 1.57 & 1.75 & 1.51 & 1.36 \\
GT/CA & 1.02 & 1.40 & 1.94 & 1.26 & 1.57 & 1.75 & 1.51 & 1.36 \\
CT/GA & 0.85 & 1.27 & 1.73 & 1.16 & 1.40 & 1.57 & 1.35 & 1.33 \\
GA/CT & 0.85 & 1.27 & 1.73 & 1.16 & 1.40 & 1.57 & 1.35 & 1.33 \\
CG/GC & 1.73 & 2.01 & 2.94 & 1.88 & 2.24 & 2.57 & 2.25 & 1.90 \\
GC/CG & 1.73 & 2.01 & 2.94 & 1.88 & 2.24 & 2.57 & 2.25 & 1.90 \\
GG/CC & 1.25 & 1.61 & 2.34 & 1.57 & 1.73 & 2.03 & 1.78 & 1.81 \\
\hline
\end{tabular}
\end{table}

\end{document}